\documentclass[a4paper,twocolumn]{article}

\usepackage[style=numeric]{biblatex}
\usepackage{hyperref}
\usepackage{graphicx}
\usepackage{microtype}
\usepackage{amsmath}
\usepackage{amssymb}
\usepackage{amsfonts}
\usepackage{amsthm}
\usepackage{mathtools}
\usepackage[capitalize]{cleveref} 
\usepackage{booktabs}
\usepackage{tikz}
\usepackage{bm}
\usepackage{multirow}
\usepackage{siunitx}
\usepackage{libertinus}

\usetikzlibrary{shapes,arrows,positioning}

\theoremstyle{plain}
\newtheorem{theorem}{Theorem}

\newtheorem{result}{Result}
\theoremstyle{definition}

\newcommand{\tran}{^\top} 
\newcommand{\rv}[1]{\bm{{#1}}} 
\newcommand{\rvec}[1]{\vec{\rv{{#1}}}} 
\newcommand{\I}{\mathbb I} 
\newcommand{\E}{\mathbb E} 
\newcommand{\dd}{\mathrm d} 
\newcommand{\nd}[2]{\frac{\dd #1}{\dd #2}}  
\newcommand{\pde}[3]{\left.\frac{\partial #1}{\partial #2}\right|_{#3}}  

\DeclareMathOperator*{\argmin}{arg\,min}
\newcommand{\CA}{\mathrm{CA}} 
\newcommand{\Norm}{\mathcal{N}} 
\DeclareMathOperator{\Beta}{B} 
\DeclareMathOperator{\Betadis}{Beta} 
\DeclareMathOperator{\Unif}{U} 
\newcommand{\rising}[1]{^{\overline{#1}}} 

\title{Probability of collision in nonlinear dynamics\\by moment propagation}
\author{
  Théo Verhelst%
  \thanks{Advanced Concepts Team, European Space Research and Technology Centre, Noordwijk, Netherlands}
  \thanks{Corresponding author, \texttt{theo.verhelst@esa.int}}
  \and
  Giacomo Acciarini%
  \thanks{Surrey Space Center / ESA ACT, European Space Research and Technology Centre, Noordwijk, Netherlands.}
  \and
  Dario Izzo\footnotemark[1]
  \and
  Francesco Biscani%
  \thanks{Skylon Dynamics, Kurpfalzstrasse 4a, 75053 Gondelsheim, Germany.}
}
\date{}

\begin{document}

\maketitle

\begin{abstract}
Estimating the probability of collision between spacecraft is crucial for risk management and collision-avoidance strategies. Current methods often rely on Gaussian assumptions and simplifications, which can be inaccurate in highly nonlinear scenarios. This paper presents a general and efficient approach for computing collision probabilities without relying on such assumptions. Using high-order multivariate Taylor polynomials, we propagate statistical moments of initial uncertainties to the point of closest approach between the spacecraft. To compute the probability of collision, we derive a semi-analytical expression for the probability density function (PDF) of the closest approach distance, inferred from the propagated moments using orthogonal polynomials. Tested on various short-term and long-term encounters in low-Earth orbit, our method accurately handles nonlinear dynamics, non-Gaussian uncertainties, and irregular distributions. This versatile framework advances space situational awareness by providing precise collision probability estimates in complex dynamical environments. Moreover, our methodology applies to any dynamical system with uncertainty in its initial state and is therefore not restricted to collision probability estimation.
\end{abstract}

\section{Introduction}
\label{sec:intro}

Estimating the probability of collision between spacecraft and space debris is the first mandatory step for managing risks in space operations and designing effective collision-avoidance strategies. With the growing density of debris in key orbital regions, a precise yet efficient approach to compute collision probabilities is needed to safeguard active satellites and space infrastructure.

Collision probability computation methods can generally be categorized into two groups: those designed for short-duration encounters, where relative motion can often be approximated as linear, and those for long-duration encounters, where nonlinear effects must be considered. Analytical approaches exist for both cases, with many relying on Gaussian uncertainty assumptions and linearized dynamics to simplify the problem~\cite{akella2000probability}. However, these assumptions can introduce significant errors in highly nonlinear scenarios, such as encounters in highly elliptical orbits, low-velocity conjunctions, or cases where uncertainty distributions deviate from Gaussianity.

The Monte Carlo (MC) approach provides a high-precision method for estimating collision probabilities by directly sampling from the uncertainty distribution and propagating many trajectories. However, MC simulations are computationally expensive, making them impractical for real-time applications, especially when extremely low-probability events must be resolved with high confidence. Recent advances, such as the Taylor Monte Carlo (TMC) method~\cite{morselli2010computing}, have sought to improve efficiency by using high-order Taylor series expansions to approximate trajectory propagation while maintaining the sampling-based nature of MC methods. While TMC can offer computational gains over standard Monte Carlo, it still requires a large number of samples to reach a significant level of statistical confidence.

Our proposed approach also employs high-order Taylor polynomials to model the flow of the dynamics. However, rather than relying on random sampling, we directly map the statistical moments of the initial conditions to the state at the closest approach, following the method presented in \cite{acciarini2024nonlinear}. These moments provide a compact representation of the uncertainty distribution at the conjunction event. The method is semi-analytical, meaning that its computational cost remains independent of the probability of collision. Additionally, the computed probability converges to the true probability as the number of moments and the order of the Taylor polynomial increase, ensuring both efficiency and accuracy.

A key challenge in our framework is translating the propagated moments into a probability density function (PDF) and ultimately into a probability of collision.  To address this, we go beyond the moment-propagation methodology of \cite{acciarini2024nonlinear} and derive an analytical expression for the PDF of the closest approach distance during an uncertain conjunction. Our method infers the PDF from the statistical moments using orthogonal polynomials~\cite{wakefield2023moment}, yielding a semi-analytical and computationally efficient solution for computing probabilities in uncertain scenarios. This process does not rely on strong simplifying assumptions (e.g., Gaussian distribution, long or short-term encounters, etc.), thereby making the method general in nature and applicable to any dynamical system, also beyond that of satellite conjunctions.

We evaluate the technique on a range of short-term and long-term encounter scenarios in low-Earth orbit, using real satellite population data from January 2022. Through these test cases, we show that our method accurately estimates the probability of collision in a number of scenarios where nonlinear dynamics and irregular distributions are prevalent, such as encounters with small relative velocities or non-Gaussian uncertainties. This is an important contribution in the field of space situational awareness, as it makes no assumption on the distribution of uncertainty or the encounter duration, unlike most current approaches. As a result, our method offers a versatile and accurate solution for collision probability estimation in complex dynamical environments.

The main contribution of this paper is a methodology to estimate the probability of collision between two spacecraft, given (a) the initial state of both satellites before the encounter, and (b) the moments of the probability distribution of the initial state. This is done in four steps:
\begin{enumerate}
    \item Compute a Taylor map of the state at the \emph{event of closest approach} between the two spacecraft, given some perturbation in the initial state. This is based on the methodology for Taylor expansion to an event manifold presented in \cite{origer2024certifying}.
    \item Propagate the moments of the square distance between the spacecraft at closest approach, using on the methodology presented in \cite{acciarini2024nonlinear}.
    \item Estimate the probability density function (PDF) of the square distance between the spacecraft at their closest approach from its moments, using the methodology presented in \cite{wakefield2023moment}.
    \item Integrate the PDF on a suitable interval to obtain the probability of collision.
\end{enumerate}

While we present our contribution in the context of estimating collision probability, its applicability extends far beyond this specific task. More generally, it can be used to estimate the probability density function of a system's state---or any function thereof---given the distribution of its initial conditions. This makes our approach a versatile tool for uncertainty quantification in dynamical systems. In particular, it provides an approximate solution to the Fokker-Planck equation without a diffusion term~\cite{risken1996fokker}, offering a computationally efficient alternative for modeling probability evolution in deterministic settings.

The remainder of this paper is organized as follows. In \cref{sec:background}, we introduce the relevant notations and concepts from probability theory. The key methods central to our contribution are described in \cref{sec:methods}, including the formalism of flow expansion with Taylor polynomials (\cref{sec:taylor_flow}) and its recent extension to event manifolds (\cref{sec:event_manifold}). The methodology for nonlinear moment propagation is detailed in \cref{sec:moment_propagation}, followed by the semi-analytical estimation of the probability density function from moments in \cref{sec:moments_to_pdf}. Our primary methodological contribution, focused on estimating the probability of collision between spacecraft, is presented in \cref{sec:contribution}. A quantitative evaluation of the proposed approach using simulated examples is provided in \cref{sec:benchmark}. Finally, we present our conclusions in \cref{sec:conclusion}.

\section{Background}
\label{sec:background}

\begin{table}
    \centering
    \begin{tabular}{ll}
        \toprule
        $P(E)$ & Probability of event $E$ \\
        $\rv x$ & Random variable \\
        $\E[\rv x]$ & Expected value of $\rv x$ \\
        $\vec x=[x^1,\dots,x^n]\tran$ & State vector \\
        $\vec x_0$ & Initial state \\
        $\vec\xi_0$ & Nominal initial state \\
        $\delta \vec x_0=\vec x_0-\vec\xi_0$ & Initial perturbation \\
        $\vec x_\CA,\vec\xi_\CA,\delta\vec x_\CA$ & State at closest approach \\
        $D_\CA$ & Distance at closest approach \\
        \bottomrule
    \end{tabular}
    \caption{Mathematical notation.}
    \label{tab:notation}
\end{table}

The mathematical notation is summarized in \cref{tab:notation}. Vector-valued functions might or might not be indicated by an arrow, depending on the context: on the one hand, multivariate functions taking some state vector $\vec x$ as input are noted $f(\vec x)=[f_1(\vec x),\dots,f_m(\vec x)]\tran$. On the other hand, we might consider a vector $\vec y$ to be a function of time, hence noted $\vec y(t)$.

The \emph{probability density function} (PDF) is a function describing the probability distribution of absolutely continuous random variables (that we call continuous random variables for short). The PDF of a continuous random variable $\rv x$ is a function $g(x)$ such that the probability that $\rv x$ belongs to some set of values $A$ is
\begin{equation}
    P(\rv x\in A)=\int_A g(x)\dd x.
\end{equation}

The \emph{expected value} of a random variable corresponds to the intuitive notion of mean, or average value. For continuous random variables, it is defined as
\begin{equation}
    \E[\rv  x]=\int x g(x)\dd x
\end{equation}
where the integral is taken over the set of all possible values that $\rv x$ can take. The expected value is a special case of the more general notion of \emph{moment}. Two types of moments exist: \emph{raw moments} and \emph{central moments}. In this paper, we are mainly concerned with the raw moments. As an abuse of language, we might refer to the raw moments simply as the moments. The raw moment of order $k\in\mathbb N$ of a continuous random variable $\rv x$ is defined as
\begin{equation}
    \E\left[\rv x^k\right]=\int x^k g(x)\dd x.
\end{equation}
One can see that the case $k=1$ corresponds to the expected value. This definition can be generalized to the multivariate case, leading to the notion of \emph{mixed raw moments}:
\begin{multline}
    \E\left[\rv x_1^{k_1}\dots \rv x_n^{k_n}\right]\\=\int\cdots\int x_1^{k_1}\dots x_n^{k_n} g(x_1,\dots,x_n)\dd x_n\dots\dd x_1
\end{multline}
where $g(x_1,\dots,x_n)$ is the joint PDF of $\rv x_1,\dots,\rv x_n$. In this paper, we adopt the notation of \emph{multi-index}, which is a tuple $\alpha=(\alpha_1,\dots,\alpha_n)\in\mathbb N^n$ such that we write, for any vector $\vec x=[x_1,\dots,x_n]\tran$,
\begin{equation}
    \vec x^\alpha=x_1^{\alpha_1}\dots x_n^{\alpha_n}.
\end{equation}
This enables the more compact notation
\begin{equation}
    \E\left[\rvec x^\alpha\right]=\int \vec x^\alpha g(\vec x)\dd\vec x.
\end{equation}
We also note $\alpha!=\alpha_1!\dots\alpha_n!$ and $|\alpha|=\alpha_1+\dots+\alpha_n$.

The task of finding a PDF that fits a given list of moments is called the \emph{moment problem} \cite{shohat1943problem}. In the continuous univariate case, it can be expressed as follows: given an infinite sequence of moments $m_1,m_2,\dots$, find a probability density function $g(x)$ such that for all $d\in\mathbb N$,
\begin{equation}
    m_d=\int_\Omega x^d g(x)\dd x
\end{equation}
for some domain of integration domain $\Omega\subseteq\mathbb R$. When $\Omega=\mathbb R$, this is called the Hamburger moment problem; when $\Omega=[0,\infty)$, it is referred to as the Stieltjes moment problem; and when $\Omega=[0,1]$, this is called the Hausdorff moment problem. Note that, for any $a,b\in\mathbb R$, the case $\Omega=[a,\infty)$ or $\Omega=(-\infty,b]$ can be reduced to the Stieltjes moment problem, and the case $\Omega=[a,b]$ can be reduced to the Hausdorff moment problem. 

\section{Methods}
\label{sec:methods}
We introduce a series of methods from the literature on dynamical systems and probability theory that will be used in our proposed approach for the estimation of collision probability.

\subsection{Taylor expansion of the flow}
\label{sec:taylor_flow}

Consider the generic dynamical system
\begin{align}
    \label{eq:def_system}
    \begin{dcases}
        \dot{\vec{x}}(t) &= f(\vec x(t), \vec q) \\
        \vec x(0) &= \vec x_0
    \end{dcases}
\end{align}
where $\vec x(t)\in\mathbb R^n$ is the state vector at time $t$, $\vec x_0\in\mathbb R^n$ is some initial condition\footnote{From now on, subscripts such as $0$, $f$ or $\CA$ applied on vectors indicate time, rather than a component of the vector. Components of state vectors are denoted with superscripts.}, $\vec q\in\mathbb R^m$ is a vector of parameters, and $f:\mathbb R^n\times\mathbb R^m\to\mathbb R^n$ represents the system dynamics. The solution to this initial value problem is represented by the \emph{flow} function, $\phi(t;\; \vec x_0, \vec q)$, with components noted $\phi_1,\dots,\phi_n$. We are interested in the variation of $\phi$ at some final time $t_f$ given some deviation $\delta\vec x_0,\delta\vec q$ in the initial conditions and parameter values. We note $\vec x_0=\vec\xi_0+\delta\vec x_0$, where $\vec\xi_0$ represents the nominal initial conditions, and $\vec q=\vec\rho+\delta\vec q$, where $\vec\rho$ represents the nominal parameter values. Let us note the nominal trajectory $\vec\xi(t)=\phi(t;\; \vec\xi_0,\vec\rho)$ and the perturbed trajectory $\vec x(t)=\phi(t;\;\vec x_0,\vec q)$. We define
\begin{align}
    \delta\vec x_f &= \vec x(t_f)-\vec\xi(t_f) \label{eq:def_delta_f_1} \\
    &= \phi(t_f;\;\vec x_0,\vec q)-\phi(t_f;\;\vec\xi_0,\vec\rho). \label{eq:def_delta_f_2}
\end{align}
For brevity, we use the notation
\begin{align}
    \vec z       &= \begin{bmatrix}\vec x_0 \\ \vec q\end{bmatrix}, &
    \delta\vec z &= \begin{bmatrix}\delta\vec x_0 \\ \delta\vec q\end{bmatrix}, &
    \vec\zeta    &= \begin{bmatrix}\vec\xi_0 \\ \vec\rho\end{bmatrix}.
\end{align}
Let us compute the Taylor polynomial expansion of order $N$ of the $i$th component of $\delta\vec x_f$ with respect to $\delta\vec z$:
\begin{align}
    &\delta x_f^i\approx\mathcal P_i(\delta\vec z) \\
    &= \sum_{|\alpha|\le N}\frac 1{\alpha!}\pde{^{|\alpha|}\phi_i(t_f;\;\vec\zeta+\delta\vec z)}{\delta\vec z^\alpha}{\delta\vec z=\vec 0}\delta\vec z^{\alpha}
\end{align}
where $\alpha=(\alpha_1,\dots,\alpha_{m+n})$ is a multi-index. We discarded the Taylor expansion of the term $\phi(t_f;\;\vec\zeta)$ in \cref{eq:def_delta_f_2} since it does not depend on $\delta\vec z$. Furthermore, since  $\vec\zeta$ and $\delta\vec z$ contribute identically to the term $\phi(t_f;\;\vec x_0,\vec q)=\phi(t_f;\;\vec\zeta+\vec\delta z)$ in \cref{eq:def_delta_f_2}, we can write
\begin{align}
    \mathcal P_i(\delta\vec z) &= \sum_{|\alpha|\le N}\frac 1{\alpha!}\pde{^{|\alpha|}\phi_i(t_f;\;\vec z)}{\vec z^\alpha}{\vec z=\vec \zeta}\delta\vec z^{\alpha}\\
    &= \sum_{|\alpha|\le N}\frac 1{\alpha!}\pde{^{|\alpha|}x^i(t_f)}{\vec z^\alpha}{\vec z=\vec\zeta}\delta\vec z^{\alpha}
\end{align}
 For simplicity, we define
\begin{equation}
    \partial_\alpha x^i(t_f)=\pde{^{|\alpha|}x^i(t_f)}{\vec z^\alpha}{\vec z=\vec\zeta}
\end{equation}
which gives the shorter notation
\begin{align}
    \delta x_f^i\approx\mathcal P_i(\delta\vec z)&=\sum_{|\alpha|\le N}\frac {\partial_\alpha x^i(t_f)}{\alpha!}\delta\vec z^{\alpha}.
\end{align}
The partial derivative terms $\partial_\alpha x^i(t_f)$ can be computed using modern integration software such as Heyoka \cite{biscani2021revisiting}.  We call the multivariate Taylor polynomial $\mathcal P=[\mathcal P_1,\dots,\mathcal P_n]$ the \emph{Taylor map}.

\subsection{Propagation to an event manifold}
\label{sec:event_manifold}

\begin{figure}
    \centering
    \begin{tikzpicture}[scale=0.8]
        \tikzset{dot/.style={
            circle,
            fill,
            minimum size=3pt,
            inner sep=0pt,
            outer sep=0pt
        }}

        
        \node[dot, label=left:$\vec\xi_0$] at (0, 0) (xi_0) {};
        \node[dot, label=left:$\vec x_0$, above right = 1 and 0.3 of xi_0] (x_0) {};
        \draw[->,>=stealth,shorten >= 1pt](xi_0) -- node[left] {$\delta\vec x_0$} (x_0);

        \node[dot, label=right:$\vec\xi_f$] at (6, 1) (xi_f) {};
        \draw[->,>=stealth,shorten >= 1pt](xi_0) .. node[below] {} controls (2.5, 1.6) and (4,0.5) .. (xi_f);
        
        \node[dot, label=above:$\vec x_f$] at (6, 1.87) (x_f) {};
        \draw[->,>=stealth,shorten >= 1pt](x_0) .. node[above] {} controls (2.2, 2.3) and (4.3, 1.7) .. (x_f);
        \draw[->,>=stealth,shorten >= 1pt](xi_f) -- node[left] {$\delta\vec x_f$} (x_f);
        
         \draw[domain=6:8, dashed, variable=\x] plot ({\x},{sqrt(\x-6) + 1});
         \draw[domain=6:8, dashed, variable=\x] plot ({\x},{-sqrt(\x-6) + 1});
         \node[right, rotate=-23] at (7, 0.4) {$e=0$};

        \node[dot, label=above:$\vec x_{\mathrm{event}}$] at (7.5, 2.22475) (x_event) {};
        \draw[->,>=stealth,shorten >= 1pt](x_f) .. controls (6.4, 1.9) .. (x_event);
    \end{tikzpicture}
    \caption{Propagation of the state to an event manifold (dashed line). The state $\vec x_f$ is reached at time $t_f$, but this final state may not lie on the event manifold anymore. The adapted Taylor map accounts for this by extending the propagation to a slightly later time, ensuring the final state lies on the manifold.}
    \label{fig:manifold}
\end{figure}
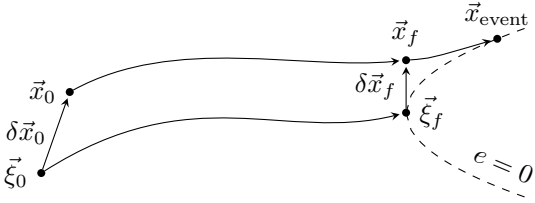

In this section, we describe the process of creating a Taylor map for the final state of a system onto an arbitrary event manifold. This approach differs from conventional Taylor maps, as presented in \cref{sec:taylor_flow}, where the final time is fixed. An event manifold refers to a subspace of the state space that satisfies a specified condition, with the time required to reach this manifold varying based on the initial state. This is illustrated in \cref{fig:manifold}. For instance, one might construct a Taylor map onto the closest approach between two spacecraft, where the time to reach this closest approach depends on the initial state. This methodology was introduced in \cite{origer2024certifying}, which considered three applications: asteroid landing, interplanetary transfer, and drone racing. The Python package Heyoka implements the necessary machinery for event detection \cite{biscani2022reliable}\footnote{A tutorial is available at \url{https://bluescarni.
github.io/heyoka.py/notebooks/map_inversion.html}}.

Formally, given an \emph{event function} $e:\mathbb R^n\to\mathbb R$, the set of values of $\vec x$ such that $e(\vec x)=a$ for some $a\in\mathbb R$ defines an \emph{event manifold} $\mathcal E$:
\begin{equation}
    \mathcal E=\{\vec x\in\mathbb R^n:e(\vec x)=a\}.
\end{equation}
Without loss of generality, we set $a=0$. The function $e$ can represent, for example, the distance to the surface of an asteroid, or, in our case, the relative velocity between two spacecraft. We assume that the nominal trajectory $\vec\xi(t)$ is such that it reaches the event manifold at time $t_f$. This is expressed mathematically as
\begin{equation}
    e\big(\vec\xi(t_f)\big)=0.
\end{equation}
To obtain a Taylor map onto this event manifold, we leverage two key ideas.

The first key idea is to add a variable to the system, noted $\varepsilon(t)$, whose value at all times is $e(\vec x(t))$. The time $t^*$ such that $\varepsilon(t^*)=0$ can vary depending on the initial condition $\vec x_0$. One can show that the time derivative of $\varepsilon$ is $\dot\varepsilon=\nabla e(\vec x(t))\cdot f$, where $\nabla e(\vec x(t))$ is the gradient of $e$ with respect to all $n$ components of $\vec x(t)$ evaluated at $\vec x(t)$, the operator $\square\cdot\square$ represents the usual dot product, and $f$ represents the dynamics of the system as in \cref{eq:def_system}. Hence the new dynamical system is
\begin{equation}
    \label{eq:dsystem_e}
    \left\{
    \begin{aligned}
        \dot{\vec{x}}(t) &= f(\vec x(t),\vec q), & \vec x(0) &= \vec x_0,\\
    \dot\varepsilon(t)&=\nabla e(\vec x(t))\cdot f, & \varepsilon(0) &= e(\vec x_0).
    \end{aligned}
    \right.
\end{equation}

The second technique is to decouple the physical final time $t_f$ from the numerical final time used by the integrator. This is achieved by introducing a formal parameter $T=t_f$ into the system (in addition to the vector $\vec{q}$) to represent the physical final time, while numerically integrating the system over a normalized time interval from $t = 0$ to $t = 1$. By treating $T$ as a formal parameter within the system rather than as an external setting for the numerical integrator, we gain the ability to manipulate it algebraically, which will prove useful in the remainder of this section. Mathematically, this is accomplished by a change of variable, letting $\tau=t/T$, and thus $\dd\tau/\dd t=1/T$. Applying the chain rule, the dynamics of the new system becomes
\begin{equation}
    \label{eq:dsystem_e_T}
    \begin{dcases}
        \nd{\vec{x}(\tau)}{\tau} &= Tf(\vec x(\tau),\vec q) \\
        \nd{\varepsilon(\tau)}{\tau}&=T\nabla e(\vec x(\tau))\cdot f
    \end{dcases}
\end{equation}
For clarity and consistency with the rest of the paper, we will use this system but still write $t$ instead of $\tau$ in the equations that follow.

After applying these two modifications, the Taylor map of the system (as derived in \cref{sec:taylor_flow}) takes two additional input arguments, a new initial state component $\delta\varepsilon_0$ and a new parameter component $\delta T$. Its output contains one additional component corresponding to $\delta\varepsilon_f$:
\begin{align}
    \label{eq:taylor_map_full_1}
    \delta x_f^1&\approx\mathcal P_1(\delta\vec x_0,\delta\varepsilon_0,\delta\vec q,\delta T) \\
    &\vdots \nonumber \\
    \label{eq:taylor_map_full_n}
    \delta x_f^n&\approx\mathcal P_n(\delta\vec x_0,\delta\varepsilon_0,\delta\vec q,\delta T) \\
    \label{eq:taylor_map_full_e}
    \delta\varepsilon_f&\approx\mathcal P_\varepsilon(\delta\vec x_0,\delta\varepsilon_0,\delta\vec q,\delta T).
\end{align}
Using only the last component of this Taylor map, we create a new map $\mathcal N$, defined as follows:
\begin{align}
    \mathcal N(\delta\vec x_0,\delta\varepsilon_0,\delta\vec q,\delta T)&=\begin{bmatrix}
        \delta\vec x_0 \\
        \delta\varepsilon_0 \\
        \delta\vec q \\
        \mathcal P_\varepsilon(\delta\vec x_0,\delta\varepsilon_0,\delta\vec q,\delta T)
    \end{bmatrix}
\end{align}
We then apply the nonlinear map inversion procedure described in \cite[Sec.~2.3.1]{berz1999modern} to obtain the inverse map $\mathcal N^{-1}$:
\begin{align}
    \mathcal N^{-1}(\delta\vec x_0,\delta\varepsilon_0,\delta\vec q,\delta\varepsilon_f)&=\begin{bmatrix}
        \delta\vec x_0 \\
        \delta\varepsilon_0 \\
        \delta\vec q \\
        \mathcal P_T(\delta\vec x_0,\delta\varepsilon_0,\delta\vec q,\delta\varepsilon_f)
    \end{bmatrix}
\end{align}
The last component of $\mathcal N^{-1}$ effectively returns the delta in time that leads to a given value of $\delta\varepsilon_f$ given some perturbations in the initial conditions. Setting as input $\delta e_f=0$ thus gives to the delta in time needed to reach the event manifold. More explicitly, for any initial perturbation, we now have
\begin{equation}
    \delta T_{\mathrm{event}}\approx\mathcal P_T(\delta\vec x_0, \delta\varepsilon_0, \delta\vec q, 0)
\end{equation}
where the value of $\delta\varepsilon_0$ is naturally computed as
\begin{equation}
    \delta\varepsilon_0 = e(\vec\xi_0+\delta\vec x_0) -e(\vec\xi_0)
\end{equation}
The state on the event manifold is computed using \crefrange{eq:taylor_map_full_1}{eq:taylor_map_full_n} a follows:
\begin{align}
    x_{\mathrm{event}}^1&=\xi^1+\delta x_{\mathrm{event}}^1 \\
    &\approx\xi^1+\mathcal P_1(\delta\vec x_0,\delta\varepsilon_0,\delta\vec q,\delta T_{\mathrm{event}}) \\
    &\vdots \nonumber \\
    x_{\mathrm{event}}^n&=\xi^n+\delta x_{\mathrm{event}}^n \\
    &\approx\xi^n+\mathcal P_n(\delta\vec x_0,\delta\varepsilon_0,\delta\vec q,\delta T_{\mathrm{event}}).
\end{align}

\subsection{Nonlinear moments propagation}
\label{sec:moment_propagation}
The Taylor map gives a way to compute the deviation of the final state from the nominal trajectory given some initial perturbation. Uncertainty propagation goes one step further by estimating the probability distribution of the final state given the probability distribution of the perturbations on the initial state. We build upon the approach in \cite{acciarini2024nonlinear} which propagates the raw moments of the distribution in nonlinear systems. We now consider $\delta\vec x_0$ to be a random variable, which we denote with a bold font $\rv\delta\rvec x_0$. The computation of the first moment (i.e., the expected value) of the perturbation at $t_f$ relies on the linearity of the expected value operator:
\begin{align}
    \E\left[\rv\delta\rv x_f^i\right] &\approx \E\left[\mathcal P_i\left(\rv\delta\rvec x_0\right)\right] \\
    &= \E\left[\sum_{|\alpha|\le N}\frac {\partial_\alpha x^i(t_f)}{\alpha!}\rv\delta\rvec x_0^{\alpha}\right]\\
    &= \sum_{|\alpha|\le N}\frac {\partial_\alpha x^i(t_f)}{\alpha!}\E\left[\rv\delta\rvec x_0^{\alpha}\right].
\end{align}
One thus needs the expression of the moments of the distribution of the initial perturbations, $\E\left[\rv\delta\rvec x_0^{\alpha}\right]$, up to order $N$. Given that the initial perturbation follows some known distribution, such as a normal or a uniform distribution, computing these moments can be done, for example, using its \emph{moment generating function}. More details can be found in \cite{acciarini2024nonlinear}.

Multivariate moments can be computed similarly, at the expanse of a larger number of operations. We only show the formula of the second mixed moment, but other moments follow a similar pattern.
\begin{align}
        &\E\left[\rv\delta\rv x_f^i\rv\delta\rv x_f^j\right] \nonumber\\
        &\approx \E\left[\sum_{|\alpha|\le N}\frac {\partial_\alpha x^i(t_f)}{\alpha!}\rv\delta\rvec x_0^{\alpha}\sum_{|\beta|\le N}\frac{\partial_\beta x^j(t_f)}{\beta!}\rv\delta\rvec x_0^{\beta}\right] \\
        &= \E\left[\sum_{|\alpha|\le N}\sum_{|\beta|\le N}\frac {\partial_\alpha x^i(t_f)}{\alpha!}\frac{\partial_\beta x^j(t_f)}{\beta!}\rv\delta\rvec x_0^{\alpha+\beta}\right]\\
        &= \sum_{|\alpha|\le N}\sum_{|\beta|\le N}\frac {\partial_\alpha x^i(t_f)}{\alpha!}\frac{\partial_\beta x^j(t_f)}{\beta!}\E\left[\rv\delta\rvec x_0^{\alpha+\beta}\right].
\end{align}

\subsection{PDF estimation from the moments}
\label{sec:moments_to_pdf}

This section describes the semi-analytical approach introduced in \cite{wakefield2023moment} to estimate a probability density function (PDF) $f(x)$ of some continuous random variable $\rv x$ when only its moments are known. This is the \emph{moment problem}, introduced in \cref{sec:background}. Numerous solutions to the moment problem have been proposed in the literature, including those based on classical orthogonal polynomial families such as Hermite polynomials~\cite{cramer1946mathematical} and Laguerre polynomials~\cite{hill1969determining,springer1979algebra}, as well as more modern techniques like kernel density functions~\cite{athanassoulis2002truncated}. The method in \cite{wakefield2023moment} generalizes the approaches based on orthogonal polynomials by establishing the general conditions for their applicability and providing a unified expression for their solutions. This framework enables the selection of the most suitable orthogonal polynomial family based on the problem's characteristics, offering flexibility and efficiency. Additionally, the method naturally extends to multidimensional problems, a feature not shared by all other approaches. However, in our application, this generalization is less relevant, as only a one-dimensional PDF is required to estimate the probability of collision, as we will see in \cref{sec:contribution}. In this section, we provide an overview of orthogonal polynomials and explain how to leverage them to estimate a univariate probability density function from its moments. We do not discuss the conditions required for the method's applicability or its generalization to the multidimensional case, which can be found in \cite{wakefield2023moment}.

The starting point of the approach is to consider a family of orthogonal polynomials, which is an infinite sequence of polynomials $P_0(x), P_1(x), P_2(x),\dots$ that are orthogonal to each other. This means that, given some \emph{weight function}\footnote{This definition using a weight function applies only when the measure considered in the inner product is absolutely continuous. This condition is always satisfied for the classical orthogonal polynomial families considered in this paper.} $w(x)$, their inner product satisfies
\begin{equation}
    \langle P_i(x),P_j(x)\rangle = \int P_i(x)P_j(x)w(x)\dd x=\delta_{i,j}
\end{equation}
where $\delta_{i,j}$ is $0$ if and only if $i\neq j$. If we further require that $\delta_{ii}=1$, then the polynomials are \emph{orthnormal}. Different weight functions give rise to different orthogonal polynomials.  From this definition, one can show that the sequence of polynomials forms a basis on the vector space of all polynomials: any polynomial $P(x)$ can be written as a unique linear combination $C_0P_0(x)+C_1P_1(x)+\dots$ Then, by the Weierestrass-Stone approximation theorem~\cite{szHokefalvi1964introduction}, any continuous function $g(x)$ on a closed interval $[a,b]$ can be approximated arbitrarily closely by polynomials. This means that there exists an infinite sequence of real coefficients $C_0,C_1,C_2,\dots$ such that
\begin{equation}
    g(x)=\sum_{i=0}^\infty C_iP_i(x).
\end{equation}

Going back to the moment problem, let us note the unknown probability density function as $f(x)$. We consider a given weight function $w(x)$, which is referred to in \cite{wakefield2023moment} as the \emph{reference distribution}. We will use both terms interchangeably. The reason behind this nomenclature will be made evident at the end of this section, where we discuss of how to find an appropriate function $w(x)$ in practice. Let us consider the expansion in terms of orthogonal polynomials $P_i(x)$ of the function $f(x)/w(x)$:
\begin{equation}
    \frac{f(x)}{w(x)}=\sum_{i=0}^\infty C_iP_i(x).
\end{equation}
If we can find the values of $C_0,C_1,\dots$, then the PDF is estimated as
\begin{align}
    \label{eq:estimated_pdf_Ci}
    f(x)=w(x)\sum_{i=0}^\infty C_iP_i(x).
\end{align}
The value of $C_i$ is in fact the inner product
\begin{align}
    \left\langle \frac{f(x)}{w(x)}, P_i(x)\right\rangle &= \int \frac{f(x)}{w(x)} P_i(x) w(x)\dd x \\
    &= \int \sum_{j=0}^\infty C_jP_j(x) P_i(x) w(x)\dd x \\
    &= \sum_{j=0}^\infty C_j\int P_j(x) P_i(x) w(x)\dd x \\
    &= \sum_{j=0}^\infty C_j \delta_{i,j} = C_i.
\end{align}
In parallel, we can develop the inner product as
\begin{align}
     \left\langle \frac{f(x)}{w(x)}, P_i(x)\right\rangle &= \int \frac{f(x)}{w(x)} P_i(x) w(x)\dd x \\
     &=  \int f(x) P_i(x) \dd x  = \E[P_i(\rv x)].
\end{align}
Therefore, we found that
\begin{align}
     C_i=\E[P_i(\rv x)]&=\E\left[\sum_{n=0}^ia_{i,n}\rv x^n\right] \\
     &=\sum_{n=0}^ia_{i,n}\E[\rv x^n]
\end{align}
where $a_{i,0},a_{i,1},\dots,a_{i,i}$ are the coefficients of the polynomial $P_i(x)$. In practice, the sum in \cref{eq:estimated_pdf_Ci} is truncated based on the number of available moments of $\rv x$.

\begin{figure}
    \centering
    \input{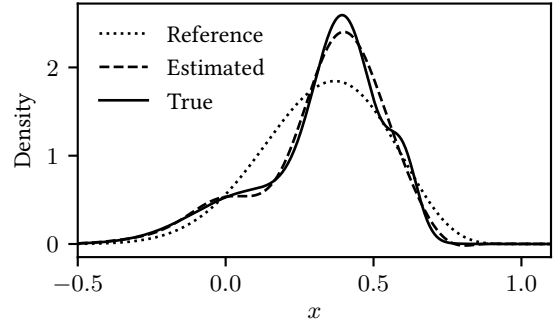}
    \caption{Example of PDF estimation, using the first 12 moments of a mixture of Gaussian distributions (solid line). The reference distribution (dotted line) is a generalized beta distribution on $[-1,1]$ with $\alpha=12.7$ and $\beta=6.4$. The estimated PDF is shown with the dashed line.}
    \label{fig:example_pdf_estimation}
\end{figure}

An important property of the reference distribution is its connection to classical orthogonal polynomial families~\cite{wakefield2023moment} For example, if $w(x)$ is chosen as the PDF of a uniform distribution, the resulting polynomials $P_0(x),P_1(x),\dots$ are in fact the Legendre polynomials (with an appropriate normalization factor). Similarly, selecting the beta distribution as $w(x)$ gives the Jacobi polynomials, the gamma distribution gives the Laguerre polynomials, and the normal distribution gives the Hermite polynomials. The appropriate weight function $w(x)$ can be selected based on prior knowledge on the distribution of $\rv x$. This weight function, also known as the reference distribution, determines the overall shape of the estimated PDF. The polynomial expansion, built on the given moments, then acts as a refinement to capture higher-order variations beyond this basic shape. This concept is illustrated in \cref{fig:example_pdf_estimation}. If a reference distribution is chosen that \emph{approximately} matches the given moments, the polynomial expansion procedure will effectively capture finer details of the true PDF. For example, the parameters of the reference distribution can be tuned to align with the first few moments of the target distribution, a technique referred to as the \emph{method of moments}. In practice, we found that matching the first two moments of the reference distribution is crucial for achieving good performance with a limited number of propagated moments.

In practice, we select the weight function as follows:
\begin{itemize}
    \item If the domain of $\rv x$ is a known interval $[u,v]$, use a beta distribution generalized to the domain $[u,v]$ (instead of the usual domain $[0,1]$), noted $\Betadis(\alpha,\beta,u,v)$, as described in \cite[Sec.~1.3.6.6.17]{heckert2002handbook}:
    \begin{equation}
        w(x)=\frac 1{\Beta(\alpha,\beta)}\frac{(x-u)^{\alpha-1}(v-x)^{\beta-1}}{(v-u)^{\alpha+\beta-1}}
    \end{equation}
    where $\Beta$ is the beta function. Note that this weight function is a generalization of the beta and uniform weight functions presented in \cite{wakefield2023moment}. We derive additional properties of this weight function necessary for PDF estimation in \cref{sec:prop_beta_weight}.
    
    \item If the domain of $\rv x$ is the half real line $[0,\infty)$, use a gamma distribution $\mathrm{Gamma}(\alpha,\lambda)$.
    \item Otherwise, use a normal distribution $\Norm(\mu,\sigma^2)$.
\end{itemize}
As discussed earlier, the parameters of these reference distributions should be fitted using the method of moments from the first moments of $\rv x$. In practice, we also observed that the gamma distribution performs poorly when most of its probability mass is ``far away'' from zero. This occurs when the shape parameter $\alpha$ of the gamma distribution is large. In this case, the normal distribution can be used instead, as the gamma distribution converges to the normal distribution for large values of $\alpha$.

\section{Probability of collision}
\label{sec:contribution}

The methods presented in \cref{sec:methods} can be applied sequentially to estimate the probability of collision between two spacecraft. Unlike common approaches such as the Akella method~\cite{akella2000probability}, our framework avoids several simplifying assumptions. First, the use of high-order Taylor polynomials eliminates the need for a linearized flow function. Second, we do not assume a specific family for the initial state probability distribution, whereas the Akella method and others typically impose a multivariate normal distribution. Third, propagating to the event manifold removes the assumption of linear trajectories at closest approach---an approximation that fails for long-term encounters, such as those in GEO~\cite{chan2004short}. Notably, our approach shares these advantages with the Taylor Monte Carlo method~\cite{morselli2010computing}, as both leverage high-order Taylor expansions on an event manifold. However, unlike Taylor Monte Carlo, our method is \emph{semi-analytical}---it does not rely on random sampling, ensuring that computation time remains constant regardless of the collision probability. We make the mild assumption of \emph{hard body radius}~\cite{nunez2022relating}, which states that the shape of both spacecraft can be simplified to two spheres of radius $R_A$ and $R_B$, such that a collision occurs whenever the relative distance between the spacecraft is less than the \emph{collision radius} $R=R_A+R_B$. 

Our methodology aims to estimate the PDF of the squared relative distance, $\rv D_\CA^2$, between two spacecraft at their closest approach. This choice, as opposed to modeling $\rv D_\CA$ directly, avoids the complications introduced by the square root operation, which can amplify errors when manipulating Taylor polynomials.  One might wonder whether integrating the PDF of the (square) distance at closest approach is indeed the same as computing the probability of collision. We demonstrate in \cref{sec:thm_distance_pc} that this is indeed the case.

The first step is to construct a Taylor polynomial $\mathcal P_{D_\CA^2}$ that approximates $\rv D_\CA^2$ as a function of perturbations in the initial state. This polynomial is derived using the identity:
\begin{align}
    \label{eq:P_D_CA_2}
    \mathcal P_{D_\CA^2}=\mathcal P_{\delta x_\CA^2}+\mathcal P_{\delta y_\CA^2}+\mathcal P_{\delta z_\CA^2}.
\end{align}
Each term on the right-hand side is obtained through the algebra of truncated Taylor polynomials~\cite{biscani2012parallel}, applied to the Taylor map of the state on the closest approach manifold, as described in \cref{sec:event_manifold}. Notably, truncated Taylor polynomials of a given order are closed under operations such as multiplication. This means that, given a Taylor polynomial $\mathcal P_{\delta x_\CA}$ of order $d$, all terms of order larger than $d$ in its square $\mathcal P_{\delta x_\CA^2}=\mathcal P_{\delta x_\CA}\cdot\mathcal P_{\delta x_\CA}$ can be safely discarded without loss of precision.

With $\mathcal P_{D_\CA^2}$ constructed, we compute its moments using the moment propagation methodology from \cref{sec:moment_propagation}, and reconstruct the PDF $f_{\rv D_\CA^2}$ with orthogonal polynomials, following \cref{sec:moments_to_pdf}. The probability of collision is then obtained by integrating $f_{\rv D_\CA^2}$ from zero to the square of the collision radius, $R^2$. We derived several closed-form expressions for this integral, depending on the choice of reference distribution, allowing the probability of collision to be computed without numerical integration. These closed-forms expression are reported in \cref{sec:closed_form_int}.

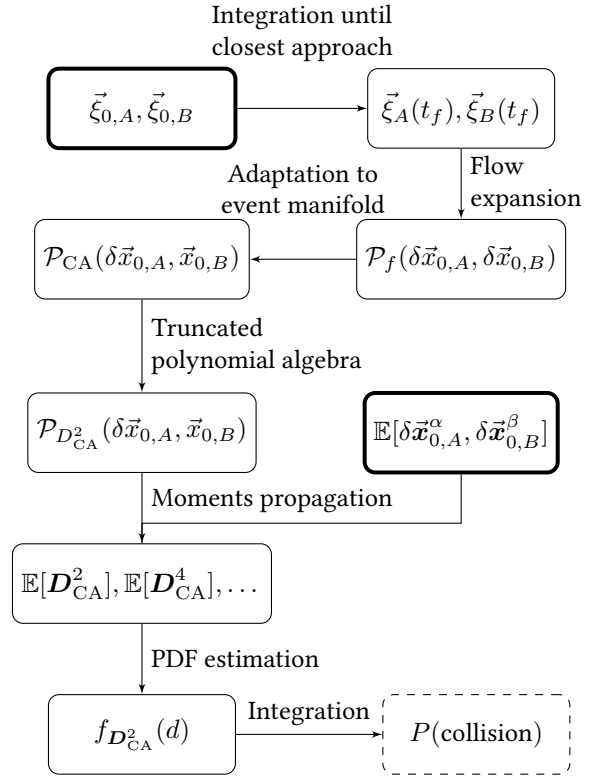
\begin{figure}
    \centering
    \begin{tikzpicture}[node distance=2cm, auto]
        \tikzstyle{block} = [
            rectangle,
            draw,
            minimum width=7em,
            text centered,
            rounded corners,
            minimum height=3em
        ]
        \tikzstyle{line} = [draw, -latex']
        \node [block,line width=1.5pt] (init) {$\vec\xi_{0,A},\vec\xi_{0,B}$};
        \node [block, right of=init, node distance = 4.2cm] (nominal) {$\vec\xi_A(t_f),\vec\xi_B(t_f)$};
        \node [block, below of=nominal] (flow_final) {$\mathcal P_f(\delta\vec x_{0,A},\delta\vec x_{0,B})$};
        \node [block, left of=flow_final, node distance = 4.2cm] (flow_event) {$\mathcal P_\CA(\delta\vec x_{0,A},\vec x_{0,B})$};
        \node [block, below of=flow_event, node distance = 2.3cm] (poly_dist) {$\mathcal P_{D_\CA^2}(\delta\vec x_{0,A},\vec x_{0,B})$};
        \node [block, right of=poly_dist, node distance=4.2cm,line width=1.5pt] (moments_0) {$\E[\delta\rvec x_{0,A}^\alpha,\delta\rvec x_{0,B}^\beta]$};
        \node [block, below of=poly_dist] (moments_dist) {$\E[\rv D_\CA^2],\E[\rv D_\CA^4],\dots$};
        \node [block, below of=moments_dist] (pdf) {$f_{\rv D_\CA^2}(d)$};
        \node [block, right of=pdf, node distance=4.4cm,dashed] (proba) {$P(\text{collision})$};
        \path [line] (init)         -- node[align=center,yshift=0.5cm] {Integration until\\closest approach} (nominal);
        \path [line] (nominal)      -- node[align=left] {Flow\\expansion} (flow_final);
        \path [line] (flow_final)   -- node[align=center,yshift=1.4cm] {Adaptation to\\event manifold} (flow_event);
        \path [line] (flow_event)   -- node[align=left] {Truncated\\polynomial algebra} (poly_dist);
        \path [line] (poly_dist)    -- node[align=center,yshift=0.1cm] {Moments propagation} (moments_dist);
        \path [draw] (moments_0) -- ++(0,-1.2) -| (moments_dist);
        \path [line] (moments_dist) -- node[align=left] {PDF estimation} (pdf);
        \path [line] (pdf)          -- node[align=left] {Integration} (proba);
    \end{tikzpicture}
    \caption{Flow diagram of our proposed approach for estimating the probability of collision. Inputs are indicated with bold borders, while the output is shown with a dashed border.}
    \label{fig:flow_diagram}
\end{figure}

A schematic representation of this procedure is depicted in \cref{fig:flow_diagram}, and can be summarized as follows:
\begin{enumerate}
    \item Consider a trajectory of a close encounter between spacecraft A and B. Specifically, integrate the dynamics from the nominal initial conditions $\vec\xi_{0,A},\vec\xi_{0,B}$ at $t_0$ up to the point of closest approach $\vec\xi_A(t_f),\vec\xi_B(t_f)$ between A and B at $t_f$ using an event detection mechanism~\cite{biscani2022reliable}.
    \item Expand the flow of the system around the nominal trajectory to obtain a Taylor map $\mathcal P_f(\delta\vec x_{0,A},\delta\vec x_{0,B})$ of the final state at $t_f$ (see \cref{sec:taylor_flow}).
    \item Form a Taylor map $\mathcal P_\CA(\delta\vec x_{0,A},\delta\vec x_{0,B})$ of the state onto the manifold of closest approach between A and B using the scheme described in \cref{sec:event_manifold}.
    \item Form a polynomial $\mathcal P_{D_\CA^2}(\delta\vec x_{0,A},\delta\vec x_{0,B})$ of the square distance between A and B at their closest approach using the algebra of truncated polynomials (Eq.~\ref{eq:P_D_CA_2})
    \item Estimate the moments $\E[\rv D_\CA^2],\E[\rv D_\CA^4],\dots$ of the square distance at closest approach using the methodology described in \cref{sec:moment_propagation}.
    \item Estimate the PDF $f_{\rv D_\CA^2}(d)$ of the distribution of the distance at closest approach using the approach described in \cref{sec:moments_to_pdf}.
    \item Integrate the PDF on the range $[0,R^2]$ where $R$ is the collision radius, to obtain the probability of collision (\cref{sec:closed_form_int}).
\end{enumerate}

While we also considered modeling the PDF of the relative position (a vector quantity) instead of the relative distance (a scalar quantity), this approach introduced significant complications due to the need for multivariate PDFs and integrals. Thus, focusing on $\rv D_\CA^2$ provides a more efficient and tractable solution.

\section{Benchmark}
\label{sec:benchmark}

To evaluate our moment-based method, we compare it against the Akella method and the Taylor Monte Carlo (TMC) method. As a ground truth reference, we use a simple Monte Carlo sampling approach. Given that we consider probabilities of collision as small as $10^{-5}$ to be significant~\cite{foster1992parametric}, the Monte Carlo and TMC approaches use $2^{23}\approx8\times 10^{6}$ random samples. We consider both a gamma and a normal reference distribution for the moment method, selecting the most effective one depending on the conjunction.

Our dataset is derived from the satellite population recorded in January 2022, using data extracted from CelesTrak. From this population, we selected satellites with semi-major axes below \num{50000} km. We then simulated their trajectories over one week using the Python package Cascade\footnote{\url{https://github.com/esa/cascade}}, assuming simple Keplerian dynamics. Conjunction events with a miss distance of less than 1000 meters were identified and recorded. From this dataset, we selected 22 encounters covering a range of miss distances and relative velocities (see \cref{tab:dataset}). Each encounter was then back-propagated by 1 hour, 1 day, or 1 week to generate different initial conditions. We model uncertainty in the initial conditions exclusively in the position components. Specifically, we consider two probability distributions: the multivariate normal distribution from \cite{pavanello2024long} (see \cref{tab:sigma_rtn}) and a uniform distribution over $[-1\mathrm m,1\mathrm m]$. Each distribution is scaled by factors of 0.1, 1, and 10, yielding six distinct uncertainty scenarios. The uncertainty distribution is initially expressed in the \emph{radial, transverse and normal} (RTN) reference frame, and then rotated to match the orientation of the spacecraft. Overall, the 22 initial encounters, 3 integration times and 6 initial distributions result in a dataset of 396 encounters scenarios.

\begin{table}[]
    \centering
    \begin{tabular}{cccc}
        \toprule
        Spacecraft & Radial & Transverse & Normal \\
        \midrule
        A & 0.625 & 10 & 3.025 \\
        B & 5.625 & 90 & 27.225 \\
         \bottomrule
    \end{tabular}
    \caption{Variance along each axis of the multivariate normal distribution, in meters squared. We assume the covariance matrix to be diagonal (cross-variable terms are zero).}
    \label{tab:sigma_rtn}
\end{table}

\sisetup{table-align-text-post=false}

\begin{table*}
    \small
    \sisetup{round-mode=places}
    \hspace{-2em}
    \begin{tabular}{
        S[table-format=2.0, round-precision=0]
        S[table-format=4.0, round-precision=0]
        S[table-format=1.1e-1, round-precision=1, exponent-mode=scientific]
        *{4}{S[table-format=1.2, round-precision=2]}
        S[table-format=4.0, round-precision=0]
        S[table-format=1.1e-1, round-precision=1, exponent-mode=scientific]
        *{4}{S[table-format=1.2, round-precision=2]}
        S[table-format=3.1, round-precision=1]
        S[table-format=4.1, round-precision=1]
    }
        \toprule
				{\multirow{2}*{ID}} & {$a_A$} & {$e_A$} & {$i_A$} & {$\Omega_A$} & {$\omega_A$} & {$E_A$ } & {$a_B$} & {$e_B$} & {$i_B$} & {$\Omega_B$} & {$\omega_B$} & {$E_B$}  & {$\|\Delta r\|$} & {$\|\Delta v\|$} \\
				& {[km]} & {[-]} & {[°]} & {[°]} & {[°]} & {[°]} & {[km]} & {[-]} & {[°]} & {[°]} & {[°]} & {[°]}  & {[m]} & {[m/s]} \\
		\midrule
		1 & 6919.677527335246 & 0.0016016783801318995 & 0.9256611736484132 & 1.2824047181922844 & 1.7724522629775499 & 3.045614960756102 & 6919.846775422072 & 0.0015816159800305996 & 0.9256346258525695 & 1.6318087625651456 & 1.5914737465132878 & 3.01507245456383 & 5.832077954898352 & 2104.7737271977057\\
		2 & 6921.2354302418125 & 0.000518406139076012 & 0.9256455306025875 & 2.5907796448561546 & 1.303464242599303 & 0.6011882353381981 & 6920.523504858145 & 0.0007472444580209027 & 0.9257806090831876 & 3.6384536579251163 & 2.355607084607068 & -1.1185805977313437 & 12.917266627676621 & 6068.83311066629\\
		3 & 6804.05801364403 & 0.0008192776815407113 & 1.6978742767350952 & 3.129149641295419 & 4.848851588067588 & 2.9838590715626534 & 6778.490188944292 & 0.005274459806297027 & 1.4432416282240346 & 0.2937819952160264 & 4.180059646720569 & -2.6247043191936115 & 16.881298360109266 & 15105.139917645241\\
		4 & 6919.39594898684 & 0.001578693142930507 & 0.9256632464142269 & 0.41644759117547736 & 1.6711385831808436 & 3.067161133218463 & 6919.440129311657 & 0.0015845988831298845 & 0.9256820541328048 & 0.5038744199938116 & 1.6369471565099003 & 3.0487286906300026 & 72.18485673793278 & 529.1961664411293\\
		5 & 6899.266691956877 & 0.0012355058226483893 & 1.2913079839101531 & 0.4914645109852071 & 2.752711218569522 & -0.7667377352272022 & 6897.9873511584265 & 0.0018346164099631373 & 1.702307861541715 & 3.320256130895726 & 2.2755119006118605 & -1.1818458623781471 & 68.87752981401985 & 14995.610825824377\\
		6 & 6726.622936887912 & 0.0008695609778404661 & 0.9287522540594129 & 5.16392849025443 & 1.4293635593363552 & 2.516032583339889 & 6726.916157148603 & 0.0008144214218901769 & 0.9287452596213696 & 5.163900263145193 & 1.4203530938245552 & 2.5251208606978697 & 169.69422900919537 & 0.5163147984061247\\
		7 & 6975.188204374477 & 0.0027649182641944726 & 1.7083529080460471 & 1.9806384176633698 & 1.3194310359612507 & -3.0156688973610817 & 6977.510066170833 & 0.0023939856287652863 & 1.7088743572222602 & 1.9848038244720159 & 1.4727216923865936 & 3.114400434643532 & 150.94459131529445 & 31.52823796176578\\
		8 & 6827.898574191971 & 0.0013001483232871645 & 1.6981900653408597 & 2.387178894393192 & 4.818695301160883 & 3.070893581951088 & 6827.222642775932 & 0.0018750985137336747 & 1.69666599629613 & 2.4273199204725335 & 5.475927858091465 & 2.417603878068304 & 161.77134097993053 & 304.16512711807627\\
		9 & 6885.679061621589 & 0.0013378755173541598 & 1.7010708271027624 & 2.1183992637463542 & 5.053520585216493 & 2.780576088519422 & 6878.809806123797 & 0.0022543637298463224 & 1.6970257628155794 & 1.9758198559064302 & 4.8264902447921605 & 2.98946360411212 & 167.21173189423163 & 1073.828953427847\\
		10 & 6919.74833523926 & 0.0004737438022318074 & 0.9256505773212883 & 1.1209303236148305 & 2.4066128141351557 & -1.0226556990602682 & 6920.062150300029 & 0.0003214385833273154 & 0.9256855582707877 & 0.5102011872197189 & 1.2420695580029155 & 0.5159505830719364 & 151.69515899917278 & 3647.2923680185945\\
		11 & 6911.252565423678 & 0.004090618941741954 & 1.7024303590802754 & 5.63668993811441 & 1.9799545786345651 & 2.7022171001180375 & 6911.104966455163 & 0.004076265498155856 & 1.7024146349173925 & 5.636531189544988 & 1.9791589677231354 & 2.7029634225298476 & 354.120438034031 & 1.2247922848011326\\
		12 & 6894.1279451581395 & 0.00046486595180563785 & 1.7018354966970874 & 2.298837692305036 & 0.6865615402212302 & 0.769519940524407 & 6893.834792393026 & 0.00046747646388073724 & 1.7017726531225716 & 2.2983181537920996 & 0.49665358305002577 & 0.9593490211455663 & 369.74249179650946 & 3.9620094564729853\\
		13 & 6892.910085377209 & 0.004031244412314593 & 1.7017499266450398 & 2.330038478497829 & 1.2695931880648108 & -2.899733815080362 & 6893.450212821153 & 0.0038304463479744665 & 1.7018109874226273 & 2.33048908912986 & 1.583365500699215 & 3.06854293336393 & 417.80751737116066 & 9.698236227183738\\
		14 & 6849.898912108429 & 0.009681874090463623 & 1.4409708204226888 & 0.3215441578217157 & 1.6905900449203273 & 2.9350359640008854 & 6863.081230269669 & 0.00974961176543684 & 1.4381395328976396 & 0.28698189259170814 & 0.8083648874868244 & -2.453224485159464 & 210.15665384481468 & 267.32306866816936\\
		15 & 6919.570653005881 & 0.0015922856640447657 & 0.9256932294862006 & 3.906588934231009 & 1.7417609258246263 & 3.0762200288554293 & 6919.431767095723 & 0.001589988053457244 & 0.9256995645965952 & 4.2558719695479486 & 1.5719877391516377 & 3.034490670686733 & 391.5047318012186 & 2104.1620112757237\\
		16 & 7858.198319790429 & 0.004520946633304077 & 1.2916006194437601 & 0.9573311450794879 & 1.5738526419822945 & -2.6246627893996353 & 7865.912300637307 & 0.003063705273496589 & 1.4398206675907381 & 3.4467963930628067 & 0.7538687342265381 & -2.8928008756349777 & 198.4796578637267 & 13166.4440817434\\
		17 & 6731.780034898822 & 0.001573919117194914 & 0.9290316842813606 & 5.161428822697904 & 1.2565168929418176 & -0.9218451176471438 & 6732.0728441904175 & 0.0015208068885380055 & 0.9290246845742248 & 5.161400696621673 & 1.2455492505752455 & -0.910883909433754 & 480.6438733411002 & 0.37623648609240995\\
		18 & 14447.097916907036 & 0.0005261923466849657 & 0.000419648037222279 & 0.1648731959337083 & 6.012028645192519 & 0.10030849825394428 & 14447.099529239586 & 0.0005193573202279659 & 6.08856486694492e-05 & 1.324001942789004 & 4.509770251841351 & 0.44330850204272826 & 996.1638909345711 & 2.223285187190043\\
		19 & 6827.9344362205375 & 0.0023368089392544495 & 1.6982249744203548 & 2.287429773827641 & 1.6153314380286075 & 3.0965245167366193 & 6828.504531070022 & 0.002304129717843955 & 1.6982199817481929 & 2.284630774998108 & 1.5939685116485367 & 3.117691977644338 & 835.3171425454262 & 21.198320651583813\\
		20 & 6896.8850517284445 & 0.002453269194595523 & 1.7010884086275768 & 2.31366904274244 & 1.4785658203153338 & -3.050866845647727 & 6894.582131217793 & 0.0029014681416263616 & 1.701096511961233 & 2.3296918237163053 & 1.6050350757658105 & 3.1075916739555836 & 849.9407988016961 & 120.51085714703487\\
		21 & 6919.670430007563 & 0.000279868332135943 & 0.9256370054505859 & 1.2954369782059394 & 1.7552149488306716 & -0.07861200746496157 & 6919.815641263383 & 0.0004046475567199201 & 0.9256428545506397 & 1.6448449453918912 & 2.3539979708264913 & -0.888713524334228 & 950.8369252781155 & 2108.682426933015\\
		22 & 6919.453458064128 & 0.00038065947915117765 & 0.9256843142177287 & 2.9542118060511724 & 1.410146486650742 & 0.320401989325445 & 6919.046857824726 & 0.0005042726431839891 & 0.9256789765716645 & 3.477970962679971 & 2.2830054139098426 & -0.8716702960502821 & 917.503991270525 & 3141.290969688321\\
        \bottomrule
    \end{tabular}
    \caption{Dataset of conjunctions used in the benchmark}
    \label{tab:dataset}
\end{table*}

\begin{figure*}
    \centering
    \includegraphics[width=16cm]{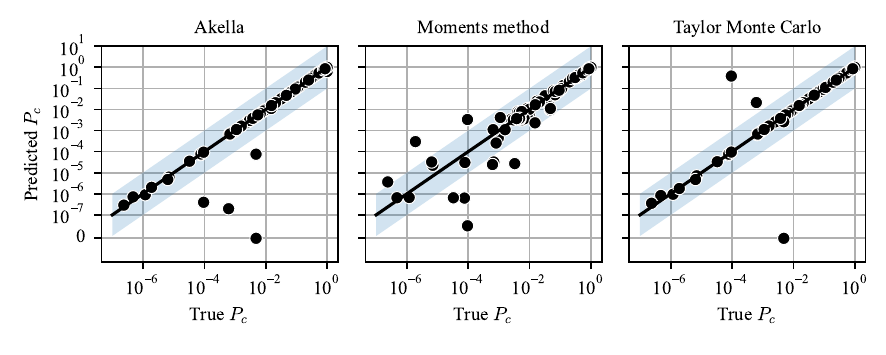}
    \caption{Comparison of true and predicted probability of collision for all three methods in the benchmark. We consider only encounters with a nonzero true probability of collision, as determined with Monte Carlo sampling. Shaded areas indicate predictions within the correct order of magnitude.}
    \label{fig:results_benchmark}
\end{figure*}

\begin{figure*}
    \centering
    \includegraphics[width=16cm]{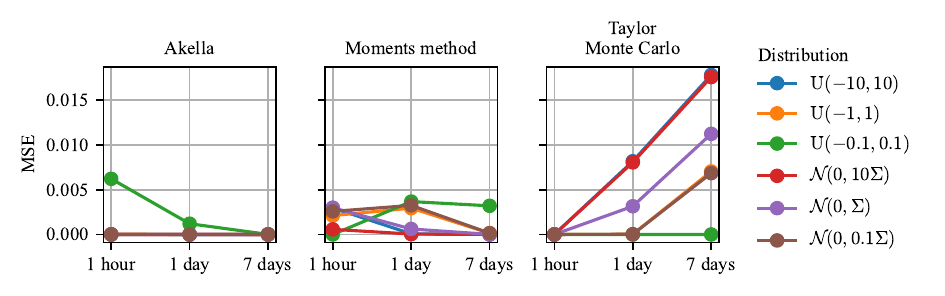}
    \caption{Mean squared error of each method as a function of integration time and the type of uncertainty distribution in the initial conditions. The error of the Taylor Monte Carlo approach increases with longer integration times, primarily due to cases where the true probability of collision is zero, but TMC predicts a nonzero probability.}
    \label{fig:plot_mse}
\end{figure*}

\begin{figure*}
    \centering
    \includegraphics[width=14cm]{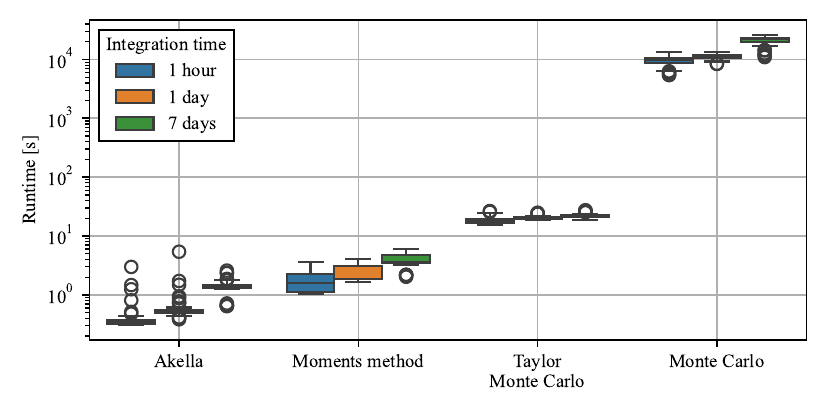}
    \caption{Boxplot of the runtime of each method, per conjunction. Different integration times shown with different colors.}
    \label{fig:runtimes}
\end{figure*}

Results are reported in \cref{fig:results_benchmark,fig:plot_mse}, and the runtime of each method is shown in \cref{fig:runtimes}. In \cref{fig:results_benchmark}, we compare the true and predicted collision probabilities for each method across all conjunctions in the dataset, considering only cases where the true probability is nonzero. We see that the Akella and Taylor Monte Carlo (TMC) methods generally provide accurate estimates, while the moments method exhibits slightly larger errors but still predicts the correct order of magnitude in most cases.

In \cref{fig:plot_mse}, we show the mean squared error of each method as a function of integration time and the type of uncertainty distribution in the initial conditions. While the Akella and moments methods maintain a relatively consistent error across integration times and distributions, the TMC approach performs worse for longer integration times. This is due to cases where the true probability of collision is zero, yet TMC predicts a nonzero probability. Although both the moment method and TMC rely on the same multivariate Taylor map, we hypothesize that the semi-analytical nature of the moment method mitigates the under-sampling issues that TMC encounters for low-probability events.

\section{Conclusion}
\label{sec:conclusion}

In this paper, we presented a novel methodology for estimating the probability of collision between two spacecraft, which is crucial for ensuring the safety and operational efficiency of space missions. Our approach leverages high-order Taylor polynomials to model the flow of spacecraft dynamics, providing a flexible and efficient means of propagating the uncertainty in initial conditions. We mapped the statistical moments of the initial state to the event of closest approach, allowing for accurate collision probability estimation without the need for random sampling, as in the Taylor Monte Carlo method.

We showed that this semi-analytical approach is general and can handle a variety of uncertainty distributions, including non-Gaussian and multivariate distributions, as well as both short- and long-term encounters. Unlike many existing methods, our approach does not rely on simplifying assumptions, making it applicable to a wide range of dynamic systems, including those with complex orbits and low relative velocities.

We compared our method to established techniques such as the Akella method and the Taylor Monte Carlo approach in a benchmark based on real-world conjunctions. The results demonstrated that our method provides a accurate estimate of collision probabilities, with computational efficiency that is independent of the probability of collision, unlike Monte Carlo-based approaches. Additionally, we highlighted the robustness of our method across different uncertainty distributions and integration times, making it well-suited for real-time space situational awareness and collision avoidance.

While our method did not outperform the conventional Akella method, it is inherently based on fewer assumptions. Future work should focus on identifying the specific regimes where failure of the assumptions underlying the Akella method provide a clear advantage to our approach. Additionally, the moment method can be refined by exploring alternative methods to compute the probability of collision from the moments of the relative distance at closest approach. While we have derived a semi-analytical solution using orthogonal polynomials, other techniques may offer different trade-offs in terms of accuracy, computational efficiency, and applicability to specific scenarios.

Additionally, the moment method is more widely applicable than just spacecraft collision analysis. It provides a versatile framework for modeling probability distributions that evolve according to any arbitrary dynamical system, using arbitrary distributions and applying them to any event manifold. This makes our approach a powerful tool for a wide range of applications beyond space situational awareness. Notably, although this paper focuses on univariate PDFs, the orthogonal polynomial method we employed is well-suited to handle multivariate distributions, enabling its application to more complex, higher-dimensional scenarios.


\section*{Disclosure statement}
AI tools have been used to improve readability, grammar, and language of this paper.

\printbibliography

\appendix

\section{Properties of the generalized beta weight function}
\label{sec:prop_beta_weight}
Let $\rv x$ be a beta-distributed random variable $\Betadis(\alpha, \beta)$. Let us consider a random variable $\rv y$ with bounded support $[u, v]$, defined as
\begin{align}
    \rv y=\rv x(v-u)+u.
\end{align}
Its PDF is
\begin{align}
    f(y)=\frac 1{\Beta(\alpha,\beta)}\frac{(y-u)^{\alpha-1}(v-y)^{\beta-1}}{(v-u)^{\alpha+\beta-1}}.
\end{align}
Its mean is
\begin{align}
    \E[\rv y]=(v-u)\E[\rv x]+u=(v-u)\frac\alpha{\alpha+\beta}+u.
\end{align}
Its second raw moment is
\begin{multline}
    \E[\rv y^2]=(v-u)^2\frac{\alpha(\alpha+1)}{(\alpha+\beta)(\alpha+\beta+1)} \\
    +2(v-u)\frac\alpha{\alpha+\beta}u+u^2.
\end{multline}
Higher raw integer moments are
\begin{equation}
    \E\left[\rv y^n\right]=\sum_{k=0}^n\binom nk(v-u)^k\frac{\alpha\rising k}{(\alpha+\beta)\rising k} u^{n-k}
\end{equation}
where $x\rising n$ is the \emph{rising factorial}, defined for $x\in\mathbb R$ and $n\in\mathbb N$ as
\begin{equation}
    x\rising n=x(x+1)(x+2)\dots(x+n-1).
\end{equation}
If we know the bounds $u,v$ and the first two raw moments $m_1=\E[\rv y]$ and $m_2=\E[\rv y^2]$, we can find the unique values of $\alpha,\beta$ that match these moments. For that, we introduce the variables $\lambda=\alpha+\beta$ and $\tau=\alpha/\lambda$. We have the equations
\begin{align}
    m_1&=(v-u)\tau+u \\
    m_2&=(v-u)^2\tau\frac{\tau\lambda+1}{\lambda+1}+2(v-u)\tau u+u^2.
\end{align}
The solution of this system of equations is
\begin{align}
    \tau    &= \frac{m_1-u}{v-u} \\
    \lambda &= \frac{(v+u)m_1-uv-m_2}{m_2-m_1^2},
\end{align}
from which we obtain
\begin{align}
    \alpha &= \tau\lambda \\
    \beta  &= (1-\tau)\lambda.
\end{align}
Following the notation in \cite{wakefield2023moment}, the functions $\sigma$ and $\tau$ for this reference distribution can be
\begin{align}
    \sigma(x) &= (x-u)(v-x) \\
    \tau(x)   &= \alpha(v-x)-\beta(x-u).
\end{align}
The function $\tau(x)$ can in fact be easily derived from $\sigma(x)$ using the identity
\begin{align}
    f(x)\tau(x)=\nd{}x(\sigma(x)f(x)).
\end{align}
The constants $B_n$ for $n\in\mathbb N$ are
\begin{align}
    B_n^2=\frac{(\alpha+\beta+2n-1)(\alpha+\beta)\rising{n-1}}{n!(v-u)^{2n}\alpha\rising n\beta\rising n}.
\end{align}

\section{The integral of the distance at closest approach is the probability of collision}
\label{sec:thm_distance_pc}

\begin{theorem}
    Let $P(\text{collision})$ be the probability that the initial position of spacecraft $A$ and $B$ lead to a closest approach with a relative distance less than the collision radius $R$. Also, let $f_{\rv D_\CA}$ be the PDF of the relative distance between the spacecraft on the closest approach event manifold. Then, the probability of collision can be computed as
    \begin{equation}
        P(\text{collision})=\int_0^R f_{\rv D_\CA}(r)\dd r.
    \end{equation}
\end{theorem}
The keep the statement of the theorem concise, the formal definition of $P(\text{collision})$ is given in the proof itself.
\begin{proof}
Let us note the state of the two spacecraft as $\vec x_A(t)=\begin{bmatrix}\vec r_A(t) \\ \vec v_A(t)\end{bmatrix}$ and $\vec x_B(t)=\begin{bmatrix}\vec r_B(t) \\ \vec v_B(t)\end{bmatrix}$. We consider the initial value problem
\begin{align}
    \label{eq:ivp_annex}
    \nd{\vec x_A(t)}t &= f(\vec x_A(t)) &
    \vec x_A(t_0)  &= \vec x_{0,A} \\
    \label{eq:ivp_annex_ic}
    \nd{\vec x_B(t)}t &= f(\vec x_B(t)) &
    \vec x_B(t_0)  &= \vec x_{0,B}.
\end{align}
We note $\phi_A(t;\; \vec x_{0,A})$ and $\phi_B(t;\; \vec x_{0,B})$ the solutions to \cref{eq:ivp_annex,eq:ivp_annex_ic}.

The set of all possible states of a single spacecraft is noted $\mathcal X$. Let $g(\vec x_{0,A}, \vec x_{0,B})$ be the function that gives the state at closest approach between the two spacecraft given their initial states:
\begin{align}
    g:\mathcal X^2\to\mathcal X^2:(\vec x_{0,A}, \vec x_{0,B})\mapsto (\vec x_A(t_\CA), \vec x_A(t_c))
\end{align}
with
\begin{align}
    t_\CA=\argmin_{t\in[0,\infty)}\,\lVert \vec r_A(t;\; \vec x_{0,A})-\vec r_B(t;\;\vec x_{0,B}, t_0) \rVert
\end{align}
where $\vec r_A(t;\; \vec x_{0,A})$ and $\vec r_B(t;\;\vec x_{0,B}, t_0)$ correspond respectively to the position components of $\phi_A(t;\; \vec x_{0,A})$ and $\phi_B(t;\; \vec x_{0,B})$.

We say that two spacecraft \emph{collide} if their position components are less than $R$ apart. Let $\mathcal C(R)\subset\mathcal X^2$ be the set of colliding states:
\begin{align}
    \mathcal C(R)=\{(\vec x_A,\vec x_B)\in\mathcal X^2: \lVert\vec r_A-\vec r_B\rVert\le R\}.
\end{align}
The probability of collision is the probability mass of the set of initial conditions that lead to a colliding state at closest approach.  For simplicity of notation, let $\vec w=(\vec x_A,\vec x_B)$ be the state of a pair of satellites, and similarly $\vec w_0=(\vec x_{0,A},\vec x_{0,B})$. Suppose that the initial state of spacecraft $A$ at time $t_0$ follows a probability density function $f_{0,A}$, and similarly the initial state of $B$ follows a probability density function $f_{0,B}$. For simplicity, we note $f_0(\vec w_0)=f_{0,A}(\vec x_{0,A})f_{0,B}(\vec x_{0,B})$ (we assume that the initial states of the two spacecraft are independently distributed). Then the probability of collision is formally defined as
\begin{align}
    P(\text{collision})=\int f_0(\vec w_0)\mathbb I[g(\vec w_0)\in\mathcal C(R)]\dd\vec w_0.
\end{align}
By the law of the unconscious statistician, we have
\begin{align}
    \int f_0(\vec w_0)\mathbb I[g(\vec w_0)\in\mathcal C(R)]\dd\vec w_0 &=\E[\mathbb I[g(\rvec w_0)\in\mathcal C(R)]] \\
    &=P(g(\rvec w_0)\in\mathcal C(R)).
\end{align}
Let $\rvec w_\CA$ be the state propagated to the closest approach, $\rvec w_\CA=g(\rvec w_0)$, and let $f_\CA$ be its PDF. By definition of the PDF,
\begin{align}
    P(g(\rvec w_0)\in\mathcal C(R))&=P(\rvec w_\CA\in\mathcal C(R)) \\
    &=\int_{\mathcal C(R)}f_\CA(\vec w_\CA)\dd\vec w_\CA.
\end{align}
Since $\mathcal C(R)$ is a subset of $\mathcal X^2$, we can write the previous expression as
\begin{align}
    &\int_{\mathcal C(R)}f_\CA(\vec w_\CA)\dd\vec w_\CA\nonumber \\
    &= \int_{\mathcal X^2} f_\CA(\vec w_\CA)\I[\vec w_\CA\in\mathcal C(R)]\dd\vec w_\CA.
\end{align}
Using Fubini's theorem,
\begin{multline}
    =\int_{\mathcal X}\int_{\mathcal X} f_\CA(\vec x_{c,A},\vec x_{c,B})\I\left[\lVert\vec r_{c,A}-\vec r_{c,B}\rVert\le R\right]\\
    \dd\vec x_{c,B}\dd\vec x_{c,A}.
\end{multline}
The integral over $\vec x_{c,B}$ can be written as an integral over a ball of radius $R$ and centered on $\vec x_{c,A}$, which we note $\mathcal B(\vec r_{c,A},R)$:
\begin{align}
    &=\int_{\mathcal X}\int_{\mathcal B(\vec r_{c,A},R)} f_\CA(\vec x_{c,A},\vec x_{c,B})\dd\vec x_{c,B}\dd\vec x_{c,A}.
\end{align}
An integral over a ball can be decomposed as a double integral over the radius $r$ and over spheres of radius $r$, noted $\mathcal S(\vec r_{c,A},r)$:
\begin{align}
    &=\int_0^R\int_{\mathcal X}\int_{\mathcal S(\vec r_{c,A},r)} f_\CA(\vec x_{c,A},\vec x_{c,B})\dd\vec x_{c,B}\dd\vec x_{c,A}\dd\vec r.
\end{align}
Note that the two inner integrals are precisely the PDF of the distance between the spacecraft:
\begin{align}
    f_{\rv D_\CA}(r)=\int_{\mathcal X}\int_{\mathcal S(\vec r_{c,A},r)} f_c(\vec x_{c,A},\vec x_{c,B})\dd\vec x_{c,B}\dd\vec x_{c,A}.
\end{align}
So, we obtained
\begin{align}
    P(\text{collision})=\int_0^R f_{\rv D_\CA}(r)\dd r.
\end{align}
\end{proof}

\section{Closed-form expressions for the integral of estimated PDFs}
\label{sec:closed_form_int}
Let us note the PDF of the square of the relative distance as $f(x)=f_{\rv D_\CA^2}(x)$. The probability of collision can be computed by integrating this PDF over the interval $[0,R^2]$. In this section, we derive closed-form expressions for this integral or representations involving well-known special functions. To extend the applicability of our results, we generalize the integral to an arbitrary domain $[a,b]$, while restricting our analysis to the one-dimensional case. Given the expansion of $f(x)$ in terms of orthogonal polynomials, the integral takes the form
\begin{align}
    \int_a^bf(x)\dd x=\int_a^bw(x)\sum_{i=0}^\infty C_iP_i(x)\dd x.
\end{align}
Since $w(x)$ is positive and $\mathcal L^2$, by Fubini we have
\begin{align}
    &\int_a^bw(x)\sum_{i=0}^\infty C_iP_i(x)\dd x\nonumber \\
    &=\sum_{i=0}^\infty C_i\int_a^bw(x)P_i(x)\dd x \\
    &=\sum_{i=0}^\infty C_i\sum_{k=0}^ia_{i,k}\int_a^bw(x)x^k\dd x. \label{eq:decomp_pdf}
\end{align}

\subsection{Uniform reference distribution}
For the uniform distribution, the weight function is
\begin{align}
    w_{\Unif(\alpha,\beta)}(x)=\frac 1{\beta-\alpha}.
\end{align}
The integral is in fact trivial:
\begin{align}
    \int_a^bw_{\Unif(\alpha,\beta)}(x)P_i(x)\dd x=\frac 1{\beta-\alpha}\int_a^b P_i(x)\dd x.
\end{align}

\subsection{Beta reference distribution}
For the beta distribution, the weight function is
\begin{align}
    w_{\Betadis(\alpha,\beta)}(x)=\frac{x^{\alpha-1}(1-x)^{\beta-1}}{\Beta(\alpha,\beta)}.
\end{align}
The integral takes the form
\begin{align}
    &\int_a^bw_{\Betadis(\alpha,\beta)}(x)P_i(x)\dd x \nonumber\\
    &=\int_a^b\frac{x^{\alpha-1}(1-x)^{\beta-1}}{\Beta(\alpha,\beta)}P_i(x)\dd x\\
    &=\sum_{k=0}^i\frac{a_{i,k}}{\Beta(\alpha,\beta)}\int_a^bx^{\alpha+k-1}(1-x)^{\beta-1}\dd x.
\end{align}
The term in the integral above corresponds to the PDF of a beta distribution with parameters $\alpha+k,\beta$ whose corresponding cumulative distribution function is the normalized incomplete beta function defined as
\begin{align}
    I_x(\alpha,\beta)=\frac 1{\Beta(\alpha,\beta)}\int_0^xt^{\alpha-1}(1-t)^{\beta-1}\dd t,
\end{align}
leading to
\begin{align}
    &\int_a^bw_{\Betadis(\alpha,\beta)}(x)P_n(x)\dd x \nonumber\\
    &=\sum_{k=0}^ia_{i,k}\frac{\Beta(\alpha+k,\beta)}{\Beta(\alpha,\beta)}\left(I_b(\alpha+k,\beta)-I_a(\alpha+k,\beta)\right).
\end{align}

\subsection{Normal reference distribution}
The situation for the normal distribution is more complicated. We leverage the following result.
\begin{result}
    Let $f$ be the probability density function of a normal random variable $\Norm(\mu,\sigma^2)$. For any real $b>0$ and integer $n\ge 0$, we have
    \begin{multline}
        \hspace{-1em}\int_0^b x^n f(x)\dd x=\sum_{k=0}^n\binom nk\mu^{n-k}\sigma^{k}\sqrt{\frac{2^{k-2}}{\pi}}\\\left(\alpha_k^n\gamma\left(\frac{k+1}2,\frac{(b-\mu)^2}{2\sigma^2}\right)-\beta_k^n\gamma\left(\frac{k+1}2,\frac{\mu^2}{2\sigma^2}\right)\right)
    \end{multline}
    where
    \begin{align}
        \alpha_k^n&=\begin{cases}
            (-1)^{k+1}&\text{if }b<\mu,\\
            1&\text{otherwise},
        \end{cases} \\
        \beta_k^n&=\begin{cases}
            (-1)^{k+1}&\text{if }\mu>0,\\
            1&\text{otherwise},
        \end{cases}
    \end{align}
    and $\gamma$ is the lower incomplete gamma function $\gamma$ defined for $s\in\mathbb C$ and $b\in\mathbb R$ as
    \begin{align}
        \gamma(s, b)=\int_0^b x^{s-1}e^{-x}\dd x.
    \end{align}
\end{result}
With this result in hand, the integral of the PDF with a normal reference distribution can be evaluated by expanding the polynomials $P_i(x)$, as in \cref{eq:decomp_pdf}. The integral on an arbitrary interval $[a,b]$ can be computed by subtracting the integral on $[0,a]$ from that on $[0,b]$.
\begin{proof}
    Let
    \begin{align}
        g_{\mu,\sigma^2}(x)=e^{-\frac{(x-\mu)^2}{2\sigma^2}}.
    \end{align}
    We want to compute $\int_0^xt^nf_{\mu,\sigma^2}(x)\dd x$. We will first prove the theorem when $\mu=0$. Let us make the substitution $x=u^2/2\sigma^2$ in the definition of $\gamma$:
    \begin{align}
        u&=\sqrt{2t\sigma^2} \\
        \dd u&=\sigma\sqrt 2\frac12t^{-\frac 12}\dd x=\frac{\sigma}{\sqrt 2}x^{-\frac 12}\dd x \\
        \frac{\sqrt 2}{\sigma}\dd u&=x^{-\frac 12}\dd x.
    \end{align}
    Substituting into the definition of $\gamma$ with $s=n$ gives
    \begin{align}
        \gamma(n,b)&=\int_0^b x^{n-\frac 12}x^{-\frac 12}e^{-x}\dd x \\
        &=\frac{\sqrt 2}{\sigma}\int_0^{\sqrt{2b\sigma^2}}\left(\frac{u^2}{2\sigma^2}\right)^{n-\frac 12}e^{-\frac{u^2}{2\sigma^2}}\dd u \\
        \gamma\left(n,\frac{b^2}{2\sigma^2}\right)&=\frac{\sqrt 2}{\sigma}\frac1{(2\sigma^2)^{n-\frac 12}}\int_0^bu^{2n-1}e^{-\frac{u^2}{2\sigma^2}}\dd u \\
        \gamma\left(\frac{n+1}2,\frac{b^2}{2\sigma^2}\right)&=\frac{\sqrt 2}{\sigma}\frac1{(2\sigma^2)^{\frac n2}}\int_0^bu^ne^{-\frac{u^2}{2\sigma^2}}\dd u \\
        \gamma\left(\frac{n+1}2,\frac{b^2}{2\sigma^2}\right)&=\frac{2^{\frac{1-n}2}}{\sigma^{n+1}}\int_0^bu^ne^{-\frac{u^2}{2\sigma^2}}\dd u.
    \end{align}
    The above equation leads to
    \begin{equation}
        \label{eq:mu_0}
        \int_0^bu^ng_{0,\sigma^2}(u)\dd u=\gamma\left(\frac{n+1}2,\frac{b^2}{2\sigma^2}\right)\sigma^{n+1}2^{\frac{n-1}2}.
    \end{equation}
    When $\mu\neq 0$, we will distinguish three cases, depending on the relative order of $\mu$, $b$ and $0$. We know that $b>0$, so either $\mu<0<b$, or $0<\mu<b$, or $0<b<\mu$ (we will deal with equalities later). For the case $\mu<0<b$, we can express the integral as
    \begin{multline}
        \int_0^bu^ng_{\mu,\sigma^2}(u)\dd u \\
        \shoveleft{=\int_\mu^bu^ng_{\mu,\sigma^2}(u)\dd u-\int_\mu^0u^ng_{\mu,\sigma^2}(u)\dd u}\\
        \shoveleft{=\int_0^{b-\mu}(x+\mu)^ng_{\mu,\sigma^2}(x+\mu)\dd x}\\
        -\int_0^{-\mu}(x+\mu)^ng_{\mu,\sigma^2}(x+\mu)\dd x
    \end{multline}
    By translation invariance, we have $g_{\mu,\sigma^2}(x+\mu)=g_{0,\sigma^2}(x)$:
    \begin{multline}
        =\int_0^{b-\mu}(x+\mu)^ng_{0,\sigma^2}(x)\dd x\\-\int_0^{-\mu}(x+\mu)^ng_{0,\sigma^2}(x)\dd x.
    \end{multline}
    Using the binomial theorem and \cref{eq:mu_0}, we obtain
    \begin{multline}
        \shoveleft{\sum_{k=0}^n\binom nk\mu^{n-k}\Bigg(\int_0^{b-\mu} x^kg_{0,\sigma^2}(x)\dd x}\\
        \shoveright{-\int_0^{-\mu} x^kg_{0,\sigma^2}(x)\dd x\Bigg)}\\
        \shoveleft{=\sum_{k=0}^n\binom nk\mu^{n-k}\sigma^{k+1}2^{\frac{k-1}2}\Bigg(\gamma\left(\frac{k+1}2,\frac{(b-\mu)^2}{2\sigma^2}\right)}\\
        -\gamma\left(\frac{k+1}2,\frac{\mu^2}{2\sigma^2}\right)\Bigg).
    \end{multline}
    That proves the case $\mu<0<b$. Now, let us look at the case $0<\mu<b$. Splitting the domain of integration at $\mu$ gives
    \begin{multline}
        \int_0^bu^ng_{\mu,\sigma^2}(u)\dd u\\
        =\int_0^\mu u^ng_{\mu,\sigma^2}(u)\dd u+\int_\mu^bu^ng_{\mu,\sigma^2}(u)\dd u.
    \end{multline}
    Let us make the substitution $x=\mu-u$ in the first term and $x=\mu+u$ in the second term:
    \begin{multline}
        \int_0^bu^ng_{\mu,\sigma^2}(u)\dd u \\
        \shoveleft{=-\int_\mu^0 (\mu-x)^ng_{\mu,\sigma^2}(\mu-x)\dd x}\\
        \shoveright{+\int_0^{b-\mu}(x+\mu)^ng_{\mu,\sigma^2}(x+\mu)\dd x}\\
        \shoveleft{=\int_0^\mu (\mu-x)^ng_{\mu,\sigma^2}(\mu-x)\dd x}\\
        \shoveright{+\int_0^{b-\mu}(x+\mu)^ng_{\mu,\sigma^2}(x+\mu)\dd u }\\
        \shoveleft{=\int_0^\mu (\mu-x)^ng_{0,\sigma^2}(x)\dd x+\int_0^{b-\mu}(x+\mu)^ng_{0,\sigma^2}(x)\dd x}.
    \end{multline}
    Expanding the terms $(x\pm\mu)^n$ gives
    \begin{multline}
        \int_0^bu^ng_{\mu,\sigma^2}(u)\dd u \\
        \shoveleft{=\sum_{k=0}^n\binom nk\mu^{n-k}\Bigg((-1)^k\int_0^\mu x^kg_{0,\sigma^2}(x)\dd x}\\
        \shoveright{+\int_0^{b-\mu}x^kg_{0,\sigma^2}(x)\dd x\Bigg) }\\
        \shoveleft{=\sum_{k=0}^n\binom nk\mu^{n-k}\sigma^{k+1}2^{\frac{k-1}2}\Bigg(\gamma\left(\frac{k+1}2,\frac{(b-\mu)^2}{2\sigma^2}\right)}\\
        +(-1)^k\gamma\left(\frac{k+1}2,\frac{\mu^2}{2\sigma^2}\right)\Bigg).
    \end{multline}
    Finally, when $0<b<\mu$, we have
    \begin{multline}
        \int_0^bu^ng_{\mu,\sigma^2}(u)\dd u=\int_0^\mu u^ng_{\mu,\sigma^2}(u)\dd u\\-\int_x^\mu u^ng_{\mu,\sigma^2}(u)\dd u.
    \end{multline}
    Let us substitute $x=\mu-u$ in both terms:
    \begin{multline*}
        =-\int_\mu^0 (\mu-x)^ng_{\mu,\sigma^2}(\mu-x)\dd x\\
        \shoveright{+\int_{\mu-b}^0 (\mu-x)^ng_{\mu,\sigma^2}(\mu-x)\dd x}\\
        \shoveleft{=\int_0^\mu (\mu-x)^ng_{\mu,\sigma^2}(\mu-x)\dd x}\\
        \shoveright{-\int_0^{\mu-b} (\mu-x)^ng_{\mu,\sigma^2}(\mu-x)\dd x} \\
        \shoveleft{=\int_0^\mu (\mu-x)^ng_{0,\sigma^2}(x)\dd x}\\
        \shoveright{-\int_0^{\mu-b} (\mu-x)^ng_{0,\sigma^2}(x)\dd x} \\
        \shoveleft{=\sum_{k=0}^n\binom nk\mu^{n-k}(-1)^k\Bigg(\int_0^\mu x^kg_{0,\sigma^2}(x)\dd x}\\
        \shoveright{-\int_0^{b-\mu}x^kg_{0,\sigma^2}(x)\dd x\Bigg)} \\
        \shoveleft{=\sum_{k=0}^n\binom nk\mu^{n-k}\sigma^{k+1}2^{\frac{k-1}2}(-1)^k}\\
        \Bigg(-\gamma\left(\frac{k+1}2,\frac{(b-\mu)^2}{2\sigma^2}\right)+\gamma\left(\frac{k+1}2,\frac{\mu^2}{2\sigma^2}\right)\Bigg).
    \end{multline*}
    By staring at which of the two incomplete gamma terms get a $-1$ in these three cases, one can verify that the definition of $\alpha^n_k$ and $\beta^n_k$ in the statement of the theorem correspond to the desired result.
\end{proof}

\end{document}